\documentclass[cits]{PoS}

\usepackage{amsmath}
\usepackage{amsfonts}
\usepackage{mathrsfs}
\usepackage{subfigure}

\newcommand{\cO}{\mathcal{O}}

\newcommand{\Tr}{{\rm Tr}}

\newcommand{\MeV}{{\rm MeV}}
\newcommand{\GeV}{{\rm GeV}}
\newcommand{\fm}{{\rm fm}}

\newcommand{\MS}{\overline{\rm MS}}

\newcommand*{\ie}{\textit{i.e.},\ }
\newcommand*{\eg}{\textit{e.g.},\ }

\newcommand{\be}{\begin{equation}}
\newcommand{\ee}{\end{equation}}
\newcommand{\bal}{\begin{align}}
\newcommand{\eal}{\end{align}}

\newcommand{\psibar}{\bar\psi}

\newcommand*{\chpt}{\raise0.4ex\hbox{$\chi$}PT}
\newcommand*{\schpt}{S\raise0.4ex\hbox{$\chi$}PT}
\newcommand*{\rschpt}{rS\raise0.4ex\hbox{$\chi$}PT}
\newcommand*{\pqchpt}{PQ\raise0.4ex\hbox{$\chi$}PT}
\newcommand*{\pqschpt}{PQ-S\raise0.4ex\hbox{$\chi$}PT}
\newcommand*{\pqrschpt}{PQ-rS\raise0.4ex\hbox{$\chi$}PT}

\def\sl#1{\rlap{\hbox{$\mskip 1 mu /$}}#1}      

\newcommand{\gammafive}{\gamma_5}
\newcommand{\gammamu}{\gamma_\mu}
\newcommand{\gammamufive}{\gamma_\mu \gamma_5}

\newcommand{\pslash}{p \!\!\!/}
\newcommand{\Asq}{\langle A^2 \rangle}

\newcommand{\mpcac}{m_{\rm PCAC}}

\title{RI/MOM renormalization constants ($\rm N_f=4$) and the strong coupling constant ($\rm N_f=2+1+1$) from twisted-mass QCD}

\ShortTitle{RI/MOM RCs and the strong coupling constant}

\author{B.~Blossier$^{a}$, Ph.~Boucaud$^{a}$, M.~Brinet$^{b}$, F.~De Soto$^{c}$,
\speaker{X.~Du}$^{b}$, M.~Gravina$^{d}$, Z.~Liu$^{e}$, V.~Morenas$^{f}$, O.~P\`ene$^{a}$, 
K.~Petrov$^{a}$, J.~Rodr\'iguez-Quintero$^{g}$\\
\llap{$^a$} Laboratoire de Physique Th\'eorique\footnote{Unit\'e Mixte de Recherche 8627 du Centre National de
la Recherche Scientifique}, CNRS et Universit\'e Paris-Sud XI, B\^atiment 210, 91405 Orsay Cedex, France\\
\llap{$^b$} Laboratoire de Physique Subatomique et de Cosmologie, CNRS/IN2P3/UJF, 53 avenue des Martyrs, 
            38026 Grenoble, France\\
\llap{$^c$} Dpto. Sistemas F\'isicos, Qu\'imicos y Naturales, Univ. Pablo de Olavide, 41013 Sevilla, Spain\\
\llap{$^d$} Cyprus? \\
\llap{$^e$} Institute of High Energy Physics, Chinese Academy of Science, Beijing 100049, China \\
\llap{$^f$} Laboratoire de Physique Corpusculaire, Universit\'e Blaise Pascal, CNRS/IN2P3, 63177 Aubi\`ere Cedex, France\\
\llap{$^g$} Dpto. F\'isica Aplicada, Fac. Ciencias Experimentales, Univ. de Huelva, 21071 Huelva, Spain\\
E-mail: \email{xdu@lpsc.in2p3.fr}}

\abstract{
We study RI/MOM renormalization constants of bilinear quark operators for $\rm N_f=4$ and the strong coupling constant 
for $\rm N_f=2+1+1$ using Wilson twisted-mass fermions. We use the ``egalitarian'' method to remove $H(4)$ hypercubic 
artifacts non-perturbatively, which enables us to study physical quantities in a wide range of momenta. We then apply 
OPE in studying the running behavior of $Z_q$ and $\alpha_s$, from which we are able to extract the Landau gauge 
dimension-two gluon condensate $\Asq$ which is of phenomenological interest.}

\FullConference{ The XXIX International Symposium on Lattice Field Theory - Lattice 2011\\
July 10-16, 2011\\
Squaw Valley, Lake Tahoe, California}

\begin{document}

\section{Introduction}
\vspace{-3mm}
The most systematic approach to study Quantum Chromodynamics (QCD) non-perturbatively is lattice QCD. By using the lattice formalism, 
one is often obliged to break some symmetries, among which is the continuum rotation symmetry. In Euclidean space, the $O(4)$ rotation 
symmetry is broken to $H(4)$ or $H(3)$ hypercubic symmetry depending on whether the lattice setup is the same on spatial and temporal 
directions. As a consequence, there are lattice artifacts which are only $H(4)$ invariant but not $O(4)$ invariant. This is 
particularly an issue for computations of quantities like renormalization constants since the associated statistical errors are often 
quite small, and the uncertainties from lattice artifacts become visible, thus deserve careful treatments. A popular solution is 
to use the ``democratic cuts'' to select data points with relatively small $H(4)$ lattice artifacts. Another approach, which we call the 
``egalitarian method''~\cite{deSoto:2007ht}, is to include the lattice artifacts explicitly in the data analysis. This approach allows 
one to use a wider range of data points and extract information from the lattice simulations more efficiently. In this work, we apply 
the ``egalitarian method'' to calculations of renormalization constants of bilinear quark operators in RI/MOM scheme and the strong coupling 
constant in MOM-type Taylor scheme, obtaining results in a wide range of momenta with lattice artifacts under control. 

On the other hand, in the low and intermediate momenta region, the perturbative running formula may not be applicable and in principle 
one should take into account non-perturbative effects. It is suggested by Operator Product Expansion (OPE) that there could be various 
contributions from condensates. In our case, as we are working in Landau gauge, we claim that there could be contributions from the 
dimension two gluon condensate $\Asq$ to some renormalization constants and running coupling constant $\alpha_s$. 
This quantity itself, $\Asq$ in Landau gauge, is of great phenomenological interests and has been studied extensively in literature (see 
~\cite{Pene:2011kg} and references therein). Here we show two independent methods to determine its value on the lattice. In both cases our results 
show strong evidences for a positive value of $\Asq$ in Landau gauge.

\section{RI/MOM renormalization constants for $\rm N_f=4$}
\vspace{-3mm}
In general, renormalization constants are needed to convert bare quantities computed on the lattice to the renormalized ones. They 
are essential for giving meaningful and accurate physical results. Renormalization constants can be calculated either perturbatively 
or non-perturbatively on the lattice. The perturbative approach suffers from systematic errors from truncations in perturbation series 
which are hard to estimate, while non-perturbative approaches suffer from discretization errors which could be systematically improved 
in practice, so generally the non-perturbative ones are preferred. In this work we adopt one particular non-perturbation method, the 
``RI'/MOM'' scheme~\cite{RIMOM}. We focus on bilinear quark operators in the form 
$\cO=\psibar\Gamma\psi$ where the Dirac matrix $\Gamma$ represents $I, \gammafive, \gammamu, \gammamufive, \gammamu\gamma_{\nu(\mu>\nu)}$.

The RI/MOM scheme is defined by requiring the amputated green function $\Lambda_\cO(p)$ equal to its tree level value at certain kinematic 
point:
    \begin{align}
    Z_q^{-1} Z_\cO \Tr[\Lambda_\cO(p) \Lambda_\cO^{tree}(p)^{-1}]_{p^2 = \mu^2} = 1, \ \ \ \ Z_q(\mu^2 = p^2) = \frac{-i}{12p^2} \Tr[S_{bare}^{-1}(p) \pslash].
    \end{align}
Here $Z_q$ is the quark wave function renormalization constant and $Z_\cO$ is the renormalization constant for operator $\cO$.
The amputated green function $\Lambda_\cO(p)$ is defined by
    \begin{align}
    S(p)&= a^4\sum_x e^{-ipx}\langle \psi(x) \psibar(0)\rangle,\ \  \Lambda_\cO (p) = S^{-1}(p) G_\cO (p) S^{-1}(p) \\
    G(p)&= a^8\sum_{x,y} e^{-ip(x-y)} \langle \psi(x) \psibar(0)\Gamma \psi(0) \psibar(y)\rangle,
    \end{align}
where the basic ingredients are quark propagators $S(p)$ and Green functions $G(p)$.

On the lattice, four-momenta are discretized, \ie they take values
\be
p_{i} = \frac{2 n_i\pi}{N_L a} \ \ n_i=-\frac{N_L}{2},\cdots,\frac{N_L}{2}, \ i=1,2,3\ \ \ \ 
p_4=\frac{(2n_4+1)\pi}{N_T a} \ \ n_4=\frac{N_T}{2},\cdots,\frac{N_T}{2}.
\ee
Note that we have used anti-periodic boundary conditions for fermions in the temporal direction. 

First we compute quark propagators and vertex functions for a wide range of momenta $n_i=-\frac{N_L}{4},\cdots,\frac{N_L}{4}, 
n_4=-\frac{N_T}{4},\cdots,\frac{N_T}{4}$. Then we apply the ``egalitarian method'' treatment to all data points to remove the hypercubic artifacts. 
The basic idea of this approach is that any polynomial function of $p$ which is invariant under $H(3)$ transformation is a polynomial function of the 
$H(3)$ invariants $p_{H3}^{[n]} \equiv \sum_{\mu=1}^3 p^n_{\mu}, \ \ n=2, 4, 6$. Here we are using $H(3)$ because we use different lengths and boundary 
conditions on temporal and spatial directions. Alternatively, one can use the functions of $H(3)$ invariants as functions of $H4$ invariants 
$p_{H4}^{[n]} \equiv \sum_{\mu=1}^4 p^n_{\mu}, \ \ n=2, 4, 6, 8$ plus the $H(4)$ to $H(3)$ symmetry breaking term $\propto (p_4^2 - {\vec{p}}^2/3)$. 
In practice, we find that the $H(4)$ to $H(3)$ breaking term is always small and can be safely ignored in the analysis. We thus use the ansatz that 
renormalization constants can be expressed as functions of $H(4)$ invariants
    \begin{align}
        Z^{latt}(ap, a^2\Lambda^2_{QCD}) &= Z^{hyp\_corr}(a^2p^2, ap_4, a^2\Lambda_{QCD}^2)
            + R(a^2p^2, a^2\Lambda^2_{QCD}) a^2 \frac{p^{[4]}}{p^2} + \cdots, \label{eq:Zqhypcorr}
    \end{align}
where $Z^{latt}$ is the raw data on the lattice and $Z^{hyp\_corr}$ is the hypercubic corrected quantity with (almost) no $H(4)$ lattice 
artifacts. Note that there could still be $O(4)$ invariant lattice artifacts in the quantity $Z^{hyp\_corr}$ and we will deal with them later.

Gauge configurations are fixed to Landau gauge for calculating quark propagators. Under this gauge, OPE suggests that the dominant 
non-perturbative contribution to $Z_q$ comes from the vacuum expectation value of $\Asq\equiv\langle A^a_\mu A^{a\mu}\rangle$. 
This term should be taken into account when we run $Z_q$ defined at different renormalization scales to some high scale, \eg 10\,\GeV, where 
perturbation theory is applicable for converting from RI/MOM scheme to $\MS$ scheme. Basically, we combine the perturbative running with OPE 
expansion in a single formula
    \begin{align}
        Z_q^{hyp\_corr}(p^2) = Z^{pert}(p^2, \mu^2) \left(1+\frac{C^Z_{Wilson}(p^2, \mu^2)}{Z^{pert}(p^2, \mu^2)}\langle A^2(\mu^2)\rangle\right) 
        + c_{a2p2}a^2 p^2,
        \label{eq:Zqrunning}
    \end{align}
where $Z^{pert}$ is the perturbative running which is available up to four-loop order~\cite{Chetyrkin:1999pq}, $C^Z_{Wilson}$ is the Wilson coefficient 
for $\Asq$ which is available up to $\cO(\alpha^4)$ in certain schemes~\cite{Chetyrkin:2009kh}. The last term in Eq.~(\ref{eq:Zqrunning}) is the 
remnant discretization artifact which is invariant under $O(4)$ symmetry and is left over from the hypercubic corrections. 

\subsection{Lattice setup}
\vspace{-1mm}
Our goal is to calculate the renormalization constants for $\rm N_f=2+1+1$ twisted-mass~\cite{Frezzotti:2000nk} ensembles generated by the European Twisted 
Mass (ETM) Collaboration. As we are working in the mass-independent scheme, in which all the renormalization constants need to be extrapolated to the 
chiral limit, the strange and charm quarks used in these configurations are too heavy for our purposes. Instead, ETMC has generated dedicated $\rm N_f=4$ 
ensembles with four degenerate (twisted-mass) light quarks. It turns out that for $\rm N_f=4$ ensembles significant amount of work is needed to tune the 
action to the maximal twisted point~\cite{David:lat10} for achieving automatic $\cO(a)$ improvement. In practice, another 
solution was used: simulating with a pair of ensembles with equal absolute values of $\mpcac$ (or $\theta$) but opposite signs, then removing the $\cO(a)$ 
artifacts by averaging the quantities obtained from these two ensembles. In this way, one avoids the work of fine tuning with the price of doubling 
the number of ensembles, and the strategy is indeed feasible as shown by David~\cite{David:lat10} in their preliminary studies.

To summarize, the procedures we need to go through for a complete analysis of the renormalization constants consist the following:
\begin{itemize}
\vspace{-1mm}
\item $\theta$-average to remove $\cO(a)$ artifacts.
\vspace{-1mm}
\item Valence chiral limit to remove possible Goldstone-pole contributions for some RCs. 
\vspace{-1mm}
\item ``Egalitarian'' hypercubic corrections to remove $H(4)$ discretization artifacts, \ie $\cO(a^2\frac{p^{[4]}}{p^2})$ terms.
\vspace{-1mm}
\item Perturbative running plus OPE to run the renormalization constants to some high energy scale, \eg 10\,\GeV. The $O(4)$ invariant lattice artifacts 
      are removed in this step.
\vspace{-1mm}
\item Sea quark chiral limit to remove remnant $\theta$-dependence~\cite{David:lat10}.
\vspace{-1mm}
\end{itemize}

\begin{figure}
	\centering
    \subfigure[Hypercubic corrections]{
        \includegraphics[scale=0.27]{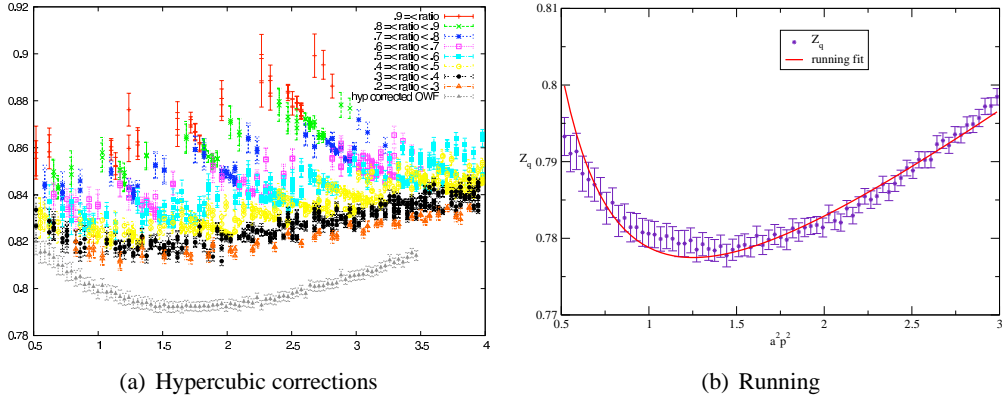}
		\label{fig:fishbone}}
	\ 
    \subfigure[Running]{
        \includegraphics[scale=0.27]{Zq_running_fit_3m3p.eps}
		\label{fig:Zqrunning}}
    \caption{The left plot shows hypercubic corrections of $Z_q$. Different colors represent data points with different values of $ratio\equiv p^{[4]}/(p^2)^2$. The yellow data
        points are those with relatively small $H(4)$ artifacts and often selected by the ``democratic'' method. The right plot shows the running of $Z_q$ according to 
        Eq.~(\protect\ref{eq:Zqrunning}).
        }
\end{figure}

Since the spacing is limited, here we only show one example of calculations of $Z_q$, leaving a complete analysis of all RCs to later publications.
We use one pair of $24^3\times 48$ ensembles with $\beta=1.95$ corresponding to lattice spacing $a\approx 0.08\,\fm$.
In Fig.~(\ref{fig:fishbone}) we show the so-called ``fishbone'' structure for raw data of $Z_q$ and hypercubic corrections with the ``egalitarian'' 
treatment. It can be clearly seen from the plot that even with the selection of ``democratic'' points the $H(4)$ artifacts in the raw data are still 
significant ($\sim5\%$) and could give misleading results without hypercubic treatments. In Fig.~(\ref{fig:Zqrunning}) we show the running fit of $Z_q(p^2)$ 
to formula in Eq.~(\ref{eq:Zqrunning}) where we have included the contributions from $\Asq$ and $O(4)$ invariant lattice artifacts. 

After converting to physical units, we obtain the values of $Z_q^{pert}$ and the gluon condensate $\Asq$ at the scale of 10\,\GeV:
\begin{align}
Z_q^{pert}(\mu=10\,\GeV) &= 0.7423\pm0.0026, \\
g^2\Asq_{CM}|_{\mu=10\,\GeV} &= 4.44\pm0.12\,\GeV^2.
\end{align}
where only statistical errors are quoted.

\section{Running coupling constant for $\rm N_f=2+1+1$}
\vspace{-3mm}
The parameter $\Lambda_{QCD}$ is a fundamental quantity of QCD and it is studied extensively in literature. Its value can be determined 
by studying the renormalized running coupling constant $\alpha_s$ from experiment or from lattice calculations. Among many methods to calculate $\alpha_s$ on the 
lattice, one category of approaches is to study the momentum behavior of Green functions. Particularly, it was shown in Ref.~\cite{Blossier:2010ky} 
that one way to study the running coupling constant is to study the ghost-ghost-gluon vertex in Landau gauge in the so-called Taylor scheme:
\begin{align}
      \alpha_T(\mu^2) \equiv \frac{g_T^2(\mu)}{4\pi} = \lim_{\Lambda\to\infty} \frac{g_0^2(\Lambda^2)}{4\pi}G(\mu^2, \Lambda^2)F^2(\mu^2,\Lambda^2),
      \label{eq:alphaT}
\end{align}
where $F(\mu^2, \Lambda^2)$ and $G(\mu^2, \Lambda^2)$ are the ghost and gluon dressing functions respectively. The advantage of this approach is that 
only two-point functions are involved thus one expects a good signal to noise ratio. This method has been successfully applied in previous $\rm N_f=0$~\cite{Boucaud:2008gn} 
and $\rm N_f=2$~\cite{Blossier:2010ky} studies of $\Lambda_{QCD}$. Here we apply the same method to $\rm N_f=2+1+1$ ensembles where we include the dynamical strange 
and charm quarks as well as two degenerate light quarks. 

\subsection{Lattice setup}
\vspace{-1mm}
We use four $\rm N_f=2+1+1$ ensembles generated by ETMC, with different choices of $\beta$, quark masses and volumes. 
The relevant parameters are listed in Table.~(\ref{tab:ensembles}). For generating these ensembles, Iwasaki action is used for the gauge part and 
Wilson twisted-mass action is used for both the degenerate light quark doublets and heavy quark doublets. The quark action is tuned to 
maximal twist by finding the value of $\kappa_{crit}$ for which the PCAC mass vanishes $\mpcac=0$. Consequently the physical (parity-even) 
quantities are automatic $\cO(a)$ improved~\cite{Frezzotti:2003ni}.

\begin{table}[ht]
\centering
\begin{tabular}{|c|c|c|c|c|c|c|}
\hline
$\beta$ & $\kappa_{\rm crit}$ & $a \mu_l$ & $a \mu_\sigma$ & $a \mu_\delta$ &
$(L/a)^3\times T/a$ & confs. \\
\hline
1.95 & 0.1612400 & 0.0035 & 0.135 & 0.170  & $32^3\times 64$ & 50 \\
     & 0.1612400 & 0.0035 &           &           & $48^3\times 96$ & 40 \\
     & 0.1612360 & 0.0055 &           &            & $32^3\times 64$ & 50 \\   
\hline
2.1 &  0.1563570 & 0.0020 & 0.120 & 0.1385 & $48^3\times 96$ & 40 \\  
\hline
\end{tabular}
\caption{Four ensembles used in our analysis. The one with $\beta=1.95$ and volume $48^3\times 96$ is only used to check finite-volume effects.}
\label{tab:ensembles}
\end{table}

\begin{figure}
	\centering
    \subfigure[Hypercubic corrections]{
        \includegraphics[scale=0.28]{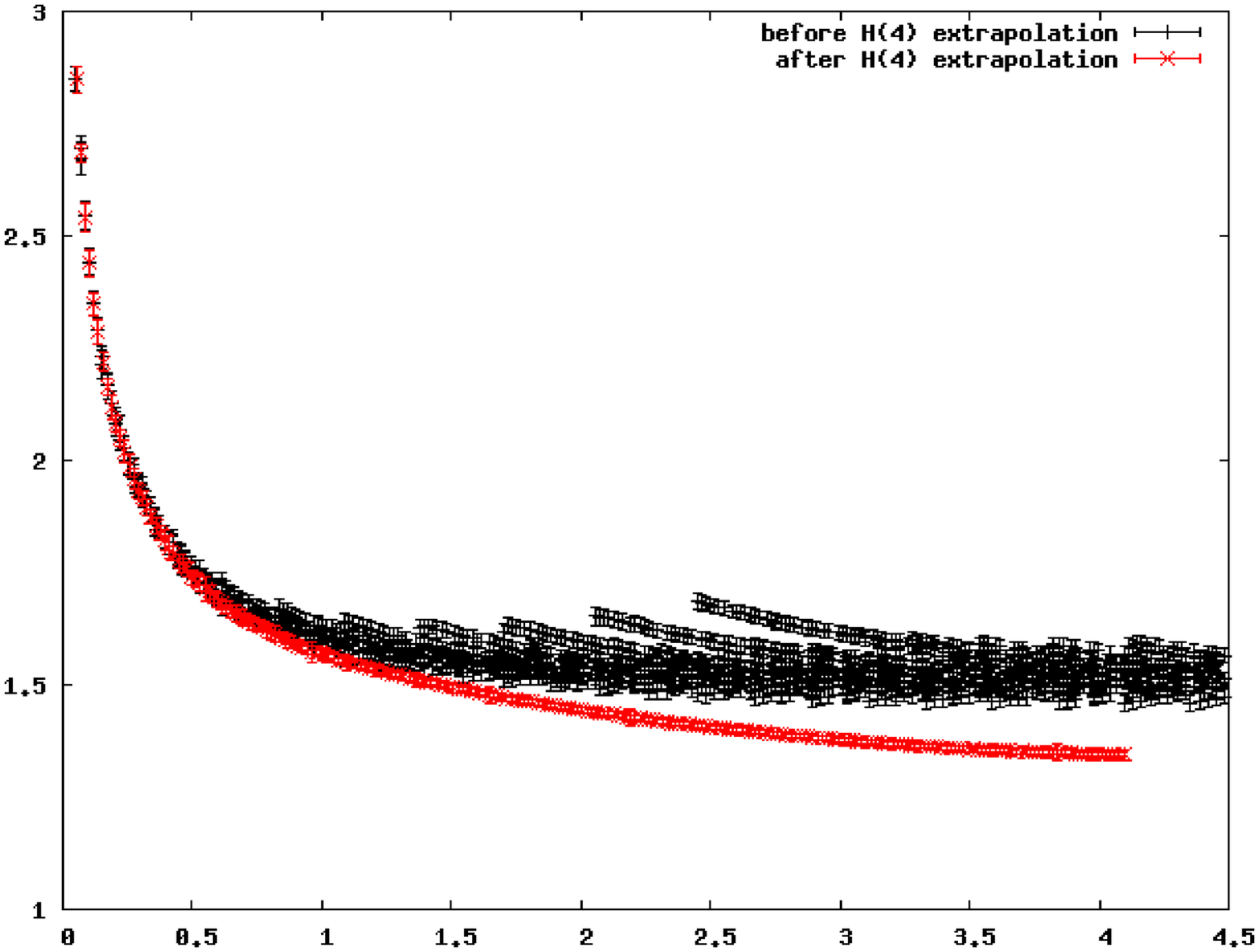}
		\label{fig:alphafishbone}}
	\ 
    \subfigure[Running]{
        \includegraphics[scale=0.28]{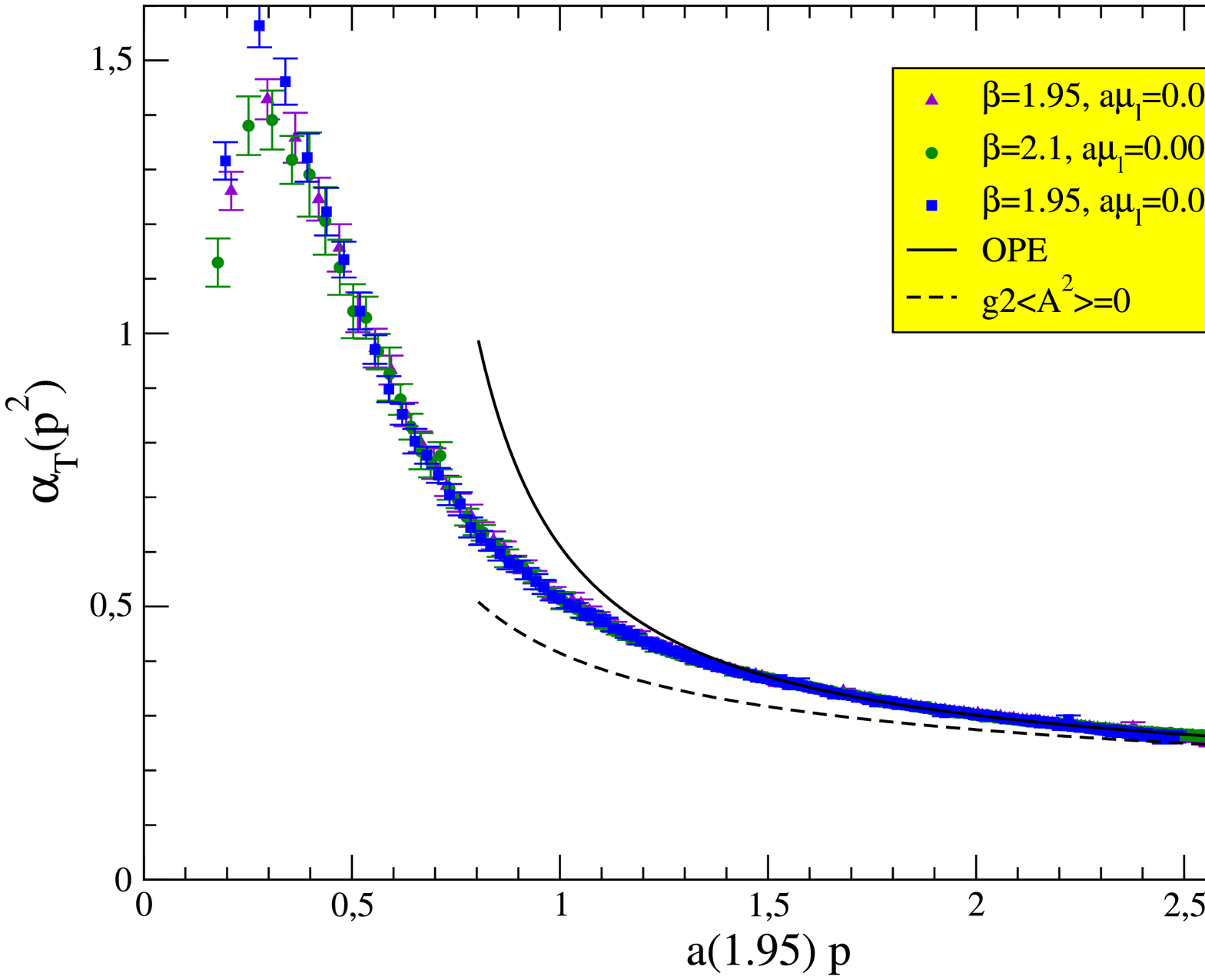}
        \label{fig:alpharunning}}
    \caption{The left plot shows hypercubic corrections of the ghost dressing function. The right plot shows the running of $\alpha_T$ according to 
    Eq.~(\protect\ref{eq:alpharunning}). }
\end{figure}

On each ensemble, we compute ghost and gluon dressing functions for a large range of momenta and apply the ``egalitarian'' hypercubic corrections to 
remove $H(4)$ artifacts. This step, shown in Fig.~(\ref{fig:alphafishbone}) for the ghost dressing function, is essential since the perturbative logarithm 
scale dependence would be missing in the large momentum region ($a^2p^2>1$) had we used the data from ``democratic'' selections. 

After hypercubic corrections, we then calculate the running coupling constant $\alpha_T$ in Taylor scheme using Eq.~(\ref{eq:alphaT}). It can be shown 
that with careful calibrations of lattice spacings~\cite{Blossier:alpha211}, the data obtained from different ensembles superimpose with each other if we 
express $\alpha_T$ in terms of momenta in units of one particular lattice spacing ($a_{\beta=1.95}$ is used here). 

Again with the argument from OPE, we claim that there could be non-perturbative contributions from a landau gauge $\Asq$ to $\alpha_T$. 
Analogous to Eq.~(\ref{eq:Zqrunning}), we fit the running coupling constant at different scales to the formula~\cite{Blossier:alpha211}
   \begin{align}
        \alpha_T(\mu^2) = \alpha_T^{pert}(\mu^2) \left(1 + \frac{9}{\mu^2} R(\mu^2, q_0^2)
										\left(\frac{\alpha_T^{pert}(\mu^2)}{\alpha_T^{pert}(q_0^2)}\right)^{\frac{27}{100}}
										\frac{g_T^2(q_0^2)\langle A^2\rangle_{R, q_0^2}}{4(N_C^2 -1)}\right)
                                     + c^{\alpha}_{a^2p^2} a^2p^2,
        \label{eq:alpharunning}
    \end{align}
where $\alpha_T^{pert}$ is available to four-loop level in perturbation theory~\cite{Chetyrkin:1999pq}. The coefficient $R(\mu^2, q_0^2)$ 
can be calculated using the Wilson coefficient which is available to $\cO(\alpha^4)$~\cite{Chetyrkin:2009kh}.

The running fit for $\alpha_T$ is shown in Fig.~(\ref{fig:alpharunning}). The solid line represents the fitted curve to the running formula Eq.~(\ref{eq:alpharunning}). 
The dotted line represents the value of $\alpha_T(\mu^2)$ from pure perturbative running if we use the PDG value $\alpha(Z^0)$ as input and run $\alpha_s$ 
to lower scales. One can clearly see the discrepancy between the pure perturbative running and the running including OPE contributions. Furthermore, a careful 
analysis~\cite{Blossier:alpha211} show that the difference between the lattice data and the fitted curve to Eq.~(\ref{eq:alpharunning}) is scaling with $1/p^6$, which seems 
to indicate that the next-to-leading correction in OPE is dominated by a $\cO(1/p^6)$ term. This is certainly a non-trivial result and needs further investigations.

By fitting to Eq.~(\ref{eq:alpharunning}), we obtain $\Lambda_{QCD}$ and $g^2\Asq$ in the Taylor scheme. Converting to $\MS$, we have 
the following results for $\beta=1.95$ lattices:
\begin{align}
\Lambda_{\MS}^{\rm N_f=4} &= 316 \pm 13\,\MeV, \ \ \ g^2\Asq_{R,\mu=10\,\GeV} = 4.5 \pm 0.4\,\GeV^2.
\end{align}

For the strong coupling constant, we can use the perturbative formula to run it to the scale of $\rm M_{Z^0}$ with matching at the bottom quark mass threshold. 
The result is $\alpha_{\MS}(\rm M_{Z^0}) = 0.1198(9)(10)$,
where the first error is statistical and the second error is the estimated uncertainty from light quark mass effects. This result agrees well with the PDG 
value $\alpha_{\MS}(\rm M_{Z^0}) = 0.1184(7)$~\cite{Nakamura:2010zzi} within errors. More details of our analysis can be found in Ref.~\cite{Blossier:alpha211}.

\section{Conclusions and outlook}
\vspace{-3mm}
In this work we studied the RI/MOM renormalization constants for bilinear quark operators at $\rm N_f=4$ and the running coupling constant at $\rm N_f=2+1+1$ using Wilson 
twisted-mass fermions. We have demonstrated that the ``egalitarian'' method is an effective way to eliminate hypercubic artifacts and its application is essential 
for obtaining reliable data in a wide range of momenta. Furthermore, we show that OPE plays an important role in the momentum range considered in our analysis. 
By taking into account the non-perturbative $\cO(1/p^2)$ contributions, we are able to extract the values of the Landau gauge gluon condensate $\Asq$ from two 
independent approaches. In both $\rm N_f=4$ and $\rm N_f=2+1+1$ cases, we obtain values of $g^2\Asq_{R,\mu=10\,\GeV}$ close to $4$ or $5\,\GeV^2$, which strongly suggest 
the existence of a non-zero Landau gauge dimension-two gluon condensate. The next step is to finish the analysis of renormalization constants for other bilinear and possibly 
twist-2 quark operators. It would also be interesting to study the strong coupling constant in $\rm N_f=4$ so that one can investigate the dependence of $\Asq$ on the 
number of sea quarks.


\end{document}